\documentclass[article,showkeys,showpacs,preprintnumbers,amsmath,amssymb,aps]{revtex4-1}
\usepackage{graphicx}
\usepackage{dcolumn}
\usepackage{bm}
\begin{document}
\preprint{APS/123-QED}
\title{Evidence of Cosmic Strings by the Observation of the Alignment of Quasar Polarization Axes on Mpc scale.}
\author{Reinoud Jan Slagter}
\email{info@asfyon.com}
\affiliation{ Asfyon, Astronomisch Fysisch Onderzoek Nederland,  1405EP Bussum \\ and \\ Department  of Physics, University of Amsterdam, The Netherlands}
\date{\today}
\begin{abstract}
The recently found alignment of the polarization axes of quasars in large quasar groups on Mpc scales, can be explained by general relativistic cosmic string networks.
By considering the cosmic string as a result of  spontaneous symmetry breaking of the gauged U(1) abelian Higgs model with topological charge $n$, many stability features of the $n$-vortex solutions of superconductivity can be taken over. Decay of the high multiplicity ($n$) super-conducting vortex into a lattice of $n$ vortices of unit magnetic flux is energetically favourable. The temporarily broken axial symmetry will leave an imprint of a preferred azimuthal-angle on the lattice. The stability of the lattice depends critically on the parameters of the model, especially when gravity comes into play.\\
In order to handle the strong nonlinear behavior of the time-dependent coupled field equations of gravity and the scalar-gauge field, we will use a high-frequency approximation scheme to second order on a warped 5D axially symmetric spacetime with the scalar-gauge field residing on the brane.\\
We consider  different winding numbers for the subsequent orders of perturbations of the scalar field. \\
A profound contribution  to the energy momentum tensor comes from the bulk spacetime and can be understand as "dark"-energy. The cosmic string becomes super-massive by  the contribution of the 5D Weyl tensor on the brane and the stored azimuthal preferences will not fade away. During the  recovery to axial symmetry, gravitational and electro-magnetic radiation will be released\\
The perturbative appearance of a non-zero energy-momentum component $T_{t\varphi}$ can be compared with the phenomenon of bifurcation along the Maclaurin-Jacobi sequence of equilibrium ellipsoids of self-gravitating compact objects, signaling the onset of secular instabilities. There is a kind of similarity  with the Goldstone-boson modes of spontaneously broken symmetries of continuous groups. The recovery of the SO(2) symmetry from the equatorial eccentricity takes place on a time-scale comparable with the emission of gravitational waves.\\
The emergent  azimuthal-angle dependency in our model can be used to explain the aligned polarization axes in large quasar groups on Mpc scales. Spin axis direction perpendicular to  the major axes of large quasar groups  when the richness decreases, can be explained as a second order effect in our approximation scheme by the higher multiplicity terms. The  preferred directions are modulo $\frac{180^o}{i}$, with $i$ an integer dependent on the $i$-th order of approximation. \\
When more data of quasars of high redshift will become available, one could proof that  the alignment emerged  after the symmetry breaking scale and must have  a cosmological  origin.
The effect of the warp factor on the second-order perturbations could also be an indication of the existence of large extra dimensions.
\end{abstract}
\maketitle
\section{\label{sec:level1}Introduction}
General relativity(GR) is by far the most successful theory constructed by theoretical physicists. One can construct in GR solutions which are related to real physical objects, for example the Kerr solution, the end stage of a collapsing spinning star. \\ A legitimate question is if there are other axially  symmetric solutions in GR.
It came as a big surprise that there exist vortex-like solutions in Einstein's theory. These vortex solutions occur as topological defects at the symmetry breaking scale in the Einstein-abelian U(1) scalar-gauge model, where the gauge field is coupled to a complex charged scalar field\cite{vil,and,ole,fel}. The solution shows a surprising resemblance with type II superconductivity of the Ginzburg-Landau(GL) theory\cite{man}, where the electro-magnetic(EM) gauge invariance is broken and the well-known Meissner effect occurs\cite{gins,abr}. One basic feature is that the lowest energy state of the scalar  field $\Phi$ is a non-zero constant with a phase freedom: $\Phi \rightarrow \Phi e^{i\varphi}$.
One says that the phase symmetry is spontaneously broken and the EM field acquires a length scale, which introduces a penetration depth of the gauge field $A_\mu$ in the superconductor and a coherence length of $\Phi$.
In the relativistic case one says that the photon acquires mass. \\
Because we have three space dimensions, these solitons behave like magnetic flux vortices ( Nielsen Olesen strings\cite{ole}) extended to tubes  and carry a quantized magnetic flux $2\pi n$, with $n$ an integer, the topological charge or winding number of the field. It was discovered by Abrikosov\cite{abr} that these vortices can form a lattice.
These localized vortices (or solitons) in the GL-theory  are  observed in experiments. The  phenomenon of magnetic flux quantization in the theory of superconductivity is characteristic for so-called ordered media.\\
The stability of these lattices depends critically on the parameters of the model, certainly when gravity comes into play. The force between the gauged vortices depends on the ratio $\alpha\equiv\frac{m_A^2}{m_\Phi^2}$, i.e., the masses of the gauge and scalar field, the GL parameter, the energy scale $\eta$ and the correlation length.
The energy of the vortex grows by increasing multiplicity $n$, so configurations with $n>1$ can be seen as multi-soliton states and it is energetically favourable for these to decay into $n$ well separated $n=1$ solitons. Vortices with high multiplicity can be formed during the symmetry breaking. The total vortex number $n$ can  be seen as the sum of multiplicities ${n_1, n_2, ..}$ of isolated points (zero's of $\Phi$)\cite{man}.

In cosmological context the confined regions of the false vacuum of $\Phi$ form a locus of trapped energy, a self-gravitating cosmic string(CS). The mass and dimension of a CS is largely determined by the energy scale at which the phase transition takes place.\\
Our universe, described by a spatially homogeneous and isotropic Friedmann Lema\^{\i}tre Robertson Walker (FLRW) spacetime, shows significant  large-scale inhomogeneous structures, for example, the cosmic web of voids with galaxies and clusters in sheets, filaments and knots, the  angular distribution in the cosmic microwave background (CMB) radiation and the recently found alignment of polarization axes of quasars in large quasar groups(LQG's) on Mpc-scales\cite{huts,tay}.
The question is if these complex nonlinear structures of deviation from isotropy and homogeneity have a cosmological origin at a moment in the early stage of the universe.
One possibility of this origin could be a CS-network formed  by the self-gravitating Einstein-scalar-gauge model. A pleasant fact is that this model has very few parameters and hence more appealing than other models such as inflationary models.
It is believed that the mass per unit length of the CS is of the order of the GUT scale, $G\mu \approx 10^{-7}$. Observational bounds, however, predict a negligible contribution of CS's to initial density perturbation from which galaxies and clusters grew. Besides the inconsistencies with the power spectrum of the CMB, radiative effects of the CS embedded in a FLRW spacetime are rapidly damped in any physical regime\cite{greg}. Further, the lensing effect of these CS's are not found yet.\\
There is, however, another possibility to detect the presence of CS's. In the framework of string theory or M-theory, super-massive CS's can be formed at a symmetry breaking scale much higher than the GUT scale, i.e., $G\mu >>1$. So their gravitational impact increases considerably, because the CS builds up a huge mass in the bulk space. Here we consider the warped brane world model of Randall-Sundrum (RS)\cite{ran1,ran2}, with one large extra dimension.
The result is that  effective 4D Kaluza-Klein(KK) modes are obtained from the perturbative  5D graviton. These KK modes will be massive from the brane viewpoint.
The modified  Einstein equations on the brane  and scalar gauge field equations will now contain contributions from the 5D Weyl tensor\cite{roy1,roy2,roy3,shir}.
In order to explore these effective field equations, we apply an approximation scheme, i.e., a multiple scale method(MSM). In this method one can handle the decay of the $n$-vortex in a perturbative way.  The MSM or high-frequency method is an approved tool to handle nonlinearities and secular terms arising in the partial differential equations(PDE) in GR.  When there is a high curvature situation, a linear approximation of the Einstein equations is not suitable\cite{choc1,choc2,slag4}.\\
In section 2 we will outline the model under consideration. This section is a revisited review of a former study\cite{slag1,slag2,slag3}. In section 3 we will explain why the correlation between the polarization axes of quasars in LQG's can be explained by cosmic string networks.
\section{\label{sec:level2}The Superconducting String Model in Warped Spacetime}
\subsection{\label{sec:level2a}The Field Equations}
We consider here  a warped five-dimensional FLRW spacetime
\begin{eqnarray}
ds^2 = {\cal W}(t,r,y)^2\Bigl[e^{2(\gamma(t,r)-\psi(t,r))}(-dt^2+ dr^2)+e^{2\psi(t,r)}dz^2+r^2 e^{-2\psi(t,r)}d\varphi^2\Bigr]+ dy^2,\label{eqn1}
\end{eqnarray}
with ${\cal W}=W_1(t,r)W_2(y)$ the warp factor. Our 4-dimensional brane is located at $y=0$. All standard model fields reside on the brane, while gravity can propagate into the bulk.
We parameterize  the self-gravitating scalar gauge field as
\begin{eqnarray}
\Phi=\eta X(t, r)e^{i n\varphi},\qquad A_\mu =\frac{1}{ \epsilon }\bigl[P(t, r)-n\bigr]\nabla_\mu\varphi, \label{eqn2}
\end{eqnarray}
with $\eta$ is the vacuum expectation value of the scalar (Higgs) field, $n$ the winding number and $\epsilon$ the gauge coupling constant.
The winding number (number of jumps in phase of the scalar field when one goes around the flux tube) is related to the quantized flux $\frac{2\pi n}{\epsilon}$ in the Ginsberg Landau theory of superconductivity (Abrikosov vortices) and the discrete values of the topological charge in the sin-Gordon theory. It seems to be strange to obtain flux quantization in a classical theory. However, there is a hidden factor $\hbar$, because the charge that appears in the Lagrangian ($\epsilon_{f}$), is not the same as the charge of the quanta of the field ($\epsilon_{p}$), i.e., $\epsilon_{p}=\hbar\epsilon_{f}$. The ansatz $A_0=0$ guarantees that static solutions are invariant
under combination of time translation and reflection $A_0({\bf x})\rightarrow-A_0({\bf x})$.
${\cal W}$ can be solved from the 5D Einstein equations\cite{slag1}
\begin{equation}
{^{5}\!G}_{\mu\nu}=-\Lambda_5{^{5}\!g}_{\mu\nu}+\kappa_5^2 \delta(y)\Bigl(-\Lambda_4 {^{4}g}_{\mu\nu}+{^{4}T}_{\mu\nu}\Bigr), \label{eqn3}
\end{equation}
with $\kappa_5= 8\pi {^{5}\!G}= 8\pi/{^{5}\!M}_{pl}^3$, $\Lambda_4$ the brane tension and $\Lambda_5$ the bulk tension.  The ${^{5}\!M}_{pl}$ is the fundamental 5D Planck mass.
The scalar-gauge field equations become\cite{garf,lag,lag2}
\begin{equation}
D^\mu D_\mu\Phi =2\frac{dV}{d\Phi^*}, \qquad{^{4}\!\nabla}^\mu F_{\nu\mu}=\frac{1}{2}i\epsilon\Bigl(\Phi(D_\nu\Phi)^*-\Phi^* D_\nu \Phi \Bigr), \label{eqn4}
\end{equation}
with $D_\mu \Phi \equiv {^{4}\!\nabla}_\mu \Phi +i\epsilon A_\mu\Phi, {^{4}\!\nabla}_\mu$ the covariant derivative with respect to ${^{4}\!g}_{\mu\nu}$, $V(\Phi)=\frac{1}{8}\beta(\Phi^2-\eta^2)^2 $ the potential of the abelian Higgs model and $\eta$ the symmetry breaking scale. $F_{\mu\nu}$ is the Maxwell tensor.
The modified Einstein equations become\cite{shir}
\begin{eqnarray}
{^{4}\!G}_{\mu\nu}=-\Lambda_{eff}{^{4}\!g}_{\mu\nu}+\kappa_4^2 {^{4}\!T}_{\mu\nu}+\kappa_5^4{\cal S}_{\mu\nu}-{\cal E}_{\mu\nu}, \label{eqn5}
\end{eqnarray}
with ${^{4}\!G}_{\mu\nu}$ the Einstein tensor calculated on the brane metric ${^{4}\!g}_{\mu\nu}= {^{5}\!g}_{\mu\nu}-n_\mu n_\nu$ and $n_\mu$ the unit vector normal to the brane.
We will consider here  $\Lambda_{eff}=0$, so we are dealing with the RS-fine tuning condition\cite{ran1}.
The last two terms on the righthand side of Eq.(\ref{eqn5}) represent the quadratic contribution of the energy-momentum tensor and the electric part of the five dimensional Weyl tensor respectively.
Is is obvious, that the cosmic string can build up a huge mass $G\mu >> 1$ by the warp factor and can induce massive KK-modes felt on the brane. The warp factor causes perturbations to be damped as they move away from the brane, so gravity looks four dimensional, at least perturbatively, to a brane world observer.
Brane world models can also explain the acceleration of the universe without the need of a cosmological constant\cite{roy2}.
Disturbances on the brane can survive the natural damping by expansion of the universe due to the warp factor. This effect was also found numerically\cite{slag1}.
\subsection{\label{sec:level2b} The Approximation Scheme}
Let us consider the formal series of the relevant fields $F_i$, i.e., the metric, the scalar field and gauge field, in a point ${\bf x}$ on a manifold M
\begin{equation}
F_i=\sum_{0}^{\infty} \frac{1}{\omega^n} F_i^{(n)}({\bf x},\xi)\label{eqn6},
\end{equation}
with $\omega>>1$ a physical expansion parameter\cite{choc1}, $\xi =\omega \Theta({\bf x})$ and $\Theta$ a scalar (phase) function on M. The small parameter $\frac{1}{\omega}$ can be the ratio of the characteristic wavelength of the perturbation to the characteristic dimension of the background or the ratio of the extra dimension y to the background dimension.
If one substitute the expansions
\begin{eqnarray}
g_{\mu\nu}=\bar g_{\mu\nu}({\bf x})+ \frac{1}{\omega}h_{\mu\nu}({\bf x},\xi)+\frac{1}{\omega^2}k_{\mu\nu}({\bf x},\xi) + ..., \cr
A_\mu=\bar A_\mu ({\bf x})+\frac{1}{\omega}B_\mu (P{\bf x},\xi) +\frac{1}{\omega^2}C_\mu ({\bf x},\xi) +... ,\cr
\Phi=\bar\Phi({\bf x}) +\frac{1}{\omega}\Psi({\bf x}, \xi)+\frac{1}{\omega^2}\Xi({\bf x}, \xi)+...,\label{eqn7}\qquad
\end{eqnarray}
into  the Einstein equations on the brane, Eq.(\ref{eqn5}), one then obtains in subsequent orders of approximation
\begin{eqnarray}
\underline {\omega^{(-1)}}:\quad
{^{4}\!G_{\mu\nu}^{(-1)}}=-{\cal E}_{\mu\nu}^{(-1)}\label{eqn8},
\end{eqnarray}
\begin{eqnarray}
\underline {\omega^{(0)}}:\quad
{^{4}\!\bar G_{\mu\nu}}+{^{4}\!G_{\mu\nu}^{(0)}}=
\kappa_4^2 \bigl({^{4}\!\bar T_{\mu\nu}}+{^{4}\!T_{\mu\nu}^{(0)}}\bigr)
+\kappa_5^4\bigl(\bar {\cal S}_{\mu\nu}+{\cal S}_{\mu\nu}^{(0)}\bigr)-\bar{\cal E}_{\mu\nu}-{\cal E}_{\mu\nu}^{(0)},\label{eqn9}
\end{eqnarray}
\begin{eqnarray}
\underline{\omega^{(1)}}:\quad
{^{4}\!G_{\mu\nu}^{(1)}}=\kappa_4^2 {^{4}\!T_{\mu\nu}^{(1)}}
+\kappa_5^4 {\cal S}_{\mu\nu}^{(1)} -{\cal E}_{\mu\nu}^{(1)}\label{eqn10}.
\end{eqnarray}
If one substitutes the expansions into the scalar-gauge field equations, one obtains to highest order $\omega^{(-1)}$ the equations
\begin{eqnarray}
l_\mu l^\mu\ddot\Psi=0 \qquad l^\mu \ddot B_\mu=0,\label{eqn11}
\end{eqnarray}
where a dot represents the derivative to the $\xi$-variable and $l_\mu$ the wave vector $l_\mu \equiv \frac{\partial \Theta}{\partial x^\mu}$. The $\omega^{(0)}$ equations for the scalar-gauge field equations will provides us the unperturbed background equations for $\bar\Phi$ and $ \bar A_\mu$ if we choose $l_\mu l^\mu=0$ from Eq.(\ref{eqn11}). After integration with respect to $\xi$ one obtains first order linear differential equations for $\dot\Psi$ and $\dot B_\mu$.
From Eq.(\ref{eqn8}) we obtain a set of restrictions on $h_{\mu\nu}$, such as the "gauge condition" $l^\alpha\bigl(\ddot h_{\alpha\nu}-\frac{1}{2}\bar g_{\alpha\nu}\ddot h\bigr)=0$.
Let us consider as a simplified case $l_\mu =[1,1,0,0,0]$. Then we let survive $h_{11}, h_{13}, h_{14}, h_{44}$ and $ h_{55} $ as independent first order perturbations of the metric. It turns out in our simplified case, that the $\omega^{(0)}$ Einstein equations provide us uncoupled background equations. They  can be solved independently.
We parameterize the scaler field in subsequent order as
\begin{eqnarray}
\bar\Phi =\eta \bar X(t,r) e^{i n_1 \varphi},\qquad \Psi = Y(t,r,\xi) e^{i n_2 \varphi},\qquad\Xi = Z(t,r,\xi) e^{i n_3 \varphi}\label{eqn12}.
\end{eqnarray}
So we break up in a perturbative way, the original vortex with winding number $n$  into  vortices with winding numbers $n_i$ with $n_{i+1}>n_i$.
In models involving more than one U(1)-charged scalar fields, the emerging strings will have log-infinity contribution to the mass and can form domain walls\cite{hill}. In our model we don't have these problems.
In the non-relativistic case the stability of the lattice of vortices will  increase when the gauge to scalar mass is $>1$\cite{bog}. In our case stability will be guaranteed  by the warp factor. Further, we parameterize $B_\mu=[B_0,B_0,0,B,0]$ and  $C_\mu=[C_0,C_0,0,C,0]$, which will fulfil the highest order perturbation equation of the gauge field, i.e., Eq.(\ref{eqn11}). So for $B_0\neq 0$ the original symmetry on the gauge field is broken, already to first order, as we shall see.
Because we are dealing with a gauge theory, this breaking of the rotational symmetry of the vortex cannot simply applied: Gauss's constraint law must be fulfilled, i.e.,
\begin{eqnarray}
\partial_\mu F^\mu_0-\frac{1}{2}i\epsilon(\Phi^ *D_0\Phi-\Phi D_0\Phi^*)=0\label{eqn13}.
\end{eqnarray}
This means in general $B_0\neq 0$.

It was found\cite{slag2,slag3} that the first order perturbations can be written as
\begin{eqnarray}
\dot{\bf U}_1=e^{\int \bar A du},\label{eqn14}
\end{eqnarray}
with $u=t-r$ and  $\bar A$ a matrix solely dependent of the background fields given by
\begin{eqnarray}
{\tiny
\bar A =
\left\vert
\begin{matrix}
               2\partial_u(\ln\bar W_1+\bar\gamma-\bar\Psi)
               & -\frac{e^{2\bar\gamma}}{r^2}(\partial_u\bar\Psi+\frac{1}{2r}) &  0
               & -\frac{1}{2}e^{2\bar\gamma-2\bar\psi}\bar W_1^2\partial_u\ln(\sqrt{r}\bar W_1)
               & 0 &  0 & \kappa_4^2\partial_u\bar X{\bf cos}(n_2-n_1\varphi) \cr
                0 &\partial_u(\ln(r\sqrt{r}\bar W_1)-2\bar\Psi) & 0 & \frac{e^{-2\bar\Psi}}{2}\bar W_1^2 r^2(\partial_u\bar\Psi+\frac{1}{2r})
                & -\kappa_4^2\frac{\partial_u\bar P}{\epsilon} & 0 & 0 \cr
                -\partial_\varphi & -\frac{e^{2\bar\gamma}}{r^2}\partial_\varphi & 2\partial_u(\ln(r\bar W_1)-\bar\Psi) & \bar W_1^2e^{2\bar\gamma -2\bar\Psi}\partial_\varphi
                & 0 & 0 & 2\kappa_4^2e^{2\bar\gamma-2\bar\Psi}\bar W_1^2\bar X\bar P{\bf sin}(n_2-n_1)\varphi \cr
                0 & 0 & 0 & 0 & 0 & 0 & 0 \cr
                0 & e^{2\bar\Psi}\frac{\partial_u\bar P}{2r^2\bar W_1^2\epsilon} & 0 & 0 & -\partial_u\bar\Psi-\frac{1}{2r} & 0 & 0 \cr
                0 & 0 & e^{2\bar\Psi}\frac{\partial_u\bar P}{\bar W_1^2 r^2\epsilon} & 0 & -\frac{e^{2\bar\gamma}}{r^2}\partial_\varphi & 0 & \epsilon e^{2\bar\gamma -2\bar\Psi}\bar W_1^2\bar X{\bf sin}(n_2-n_1)\varphi \cr
                0 & 0 & 0 & 0 & 0 & 0 & -\partial_u\ln(\frac{\bar W_1}{\sqrt{r}}) \cr
                \label{eqn15}
\end{matrix}
\right\vert  }
\end{eqnarray}
and $\dot{\bf U}_1=[\dot h_{11},\dot h_{44},\dot h_{14},\dot h_{55},\dot B,\dot B0,\dot Y]$. Note that $\dot h_{44}$ interacts with the gauge field perturbation $\dot B$, even when $\dot\Psi$ is absent.
Further,  when we should consider all first-order perturbations independent of the azimuthal-angle $\varphi$, then their still appears $\varphi$-dependent terms
${\bf sin}(n_2-n_1)\varphi$ and $ {\bf cos}(n_2-n_1)\varphi$ terms. So the deviation from axially symmetry is emergent due to the fact that $n_2 >n_1$. This becomes also clear by the non-zero off-diagonal energy-momentum tensor to  second order
\begin{eqnarray}
{^{4}\bar T}_{t\varphi}=0,\label{eqn16}
\end{eqnarray}
\begin{eqnarray}
{^{4}T}_{t\varphi}^{(0)}=\bar X\bar P\dot Y{\bf sin}[(n_2-n_1)\varphi],\label{eqn17}
\end{eqnarray}
\begin{eqnarray}
{^{4}T}_{t\varphi}^{(1)}=\Bigl[\partial_t\bar X Y(n_1-n_2-\bar P)+\bar X(\bar P\partial_t Y+\epsilon B\dot Y)\Bigr]{\bf sin}[(n_2-n_1)\varphi]+\bar X\bar P\dot Z{\bf sin}[(n_3-n_1)\varphi]\cr
+\frac{e^{2\bar\psi-2\bar\gamma}}{\bar W_1^2}\dot Y h_{14}(\partial_t\bar X-\partial_r\bar X){\bf cos}[(n_2-n_1)\varphi]+\bar X^2\bar P\epsilon B_0-\frac{1}{8}\beta(\bar X^2-\eta^2)^2h_{14}\cr
-\frac{e^{2\bar\psi-2\bar\gamma}}{2\bar W_1^2}h_{14}\Bigl[\frac{e^{2\bar\psi}}{r^2\bar W_1^2\epsilon^2}(
\partial_t\bar P-\partial_r\bar P)^2+\partial_r\bar X^2-\partial_t\bar X^2
+e^{2\bar\gamma}\frac{\bar X^2\bar P^2}{r^2}\Bigr]\label{eqn18}.
\end{eqnarray}
The azimuthal-angle dependency in ${^{4}T}_{t\varphi}^{(0)}$ is evident.
It is remarkable that in the second order ${^{4}T}_{t\varphi}^{(1)}$ there appears, even for $h_{14}=0$, two trigonometric functions with period dependent on $(n_2-n_1)$ and $(n_3-n_1)$. So as second-order effect there will be two preferred directions $mod(\pi)$.
This effect could be tested in the observed quasar polarization axes alignment, with two perpendicular preferred orientations. See section 3

The first order  differential equations for the second order perturbations  are no longer linear. If we define $\dot{\bf U}_2=[\dot k_{ij},\dot Z,\dot C ,\dot C_0]$, we have
\begin{eqnarray}
n^i\partial_i\dot{\bf U}_2=\bar D_1\dot{\bf U}_2 +D_2\dot{\bf U}_1+D_3,\label{eqn19}
\end{eqnarray}
with $\bar D_1$ a matrix solely dependent of the background fields and  $D_2, D_3$ matrices  in first order perturbations and background fields\cite{slag3}.
The righthand side will now contain terms like ${\bf cos}(n_3-n_1)\varphi$.
It turns out that these second order equations Eq.(\ref{eqn19}) contain second order derivative terms for the first order perturbations. So by imposing suitable constraints on $ \dot{\bf U}_2$, one could also solve  a system of second order PDE's for ${\bf U}_1$. If one pushes the approximation to higher orders, then one obtains in the same way second order PDE's for ${\bf U_2}$.
It must be noted that in the non-general relativistic model one obtains first order differential equations only in the Bogomol'nyi limit $m_A=m_\Phi$. In the multiple-scale approximation it is a genuine feature independent of constraints on the parameters\cite{choc1}.
\subsection{\label{sec:level2c} Excitation of Vortices}
Vortices in type-II superconductivity was first described by Abrikosov\cite{abr}. The electro-magnetic gauge invariance is spontaneously broken. The photon acquires a mass and the well-known Meissner effect occurs. The applied magnetic field penetrates into the structure in the form of quantized flux. It is the circulating supercurrents in the soft core coherence length($\zeta$) $< r < $ penetration length ($\nu$) which prevent the magnetic field from being spread out. See Figure 1.
\begin{figure}[h]
\centerline{
\includegraphics[width=4.5cm]{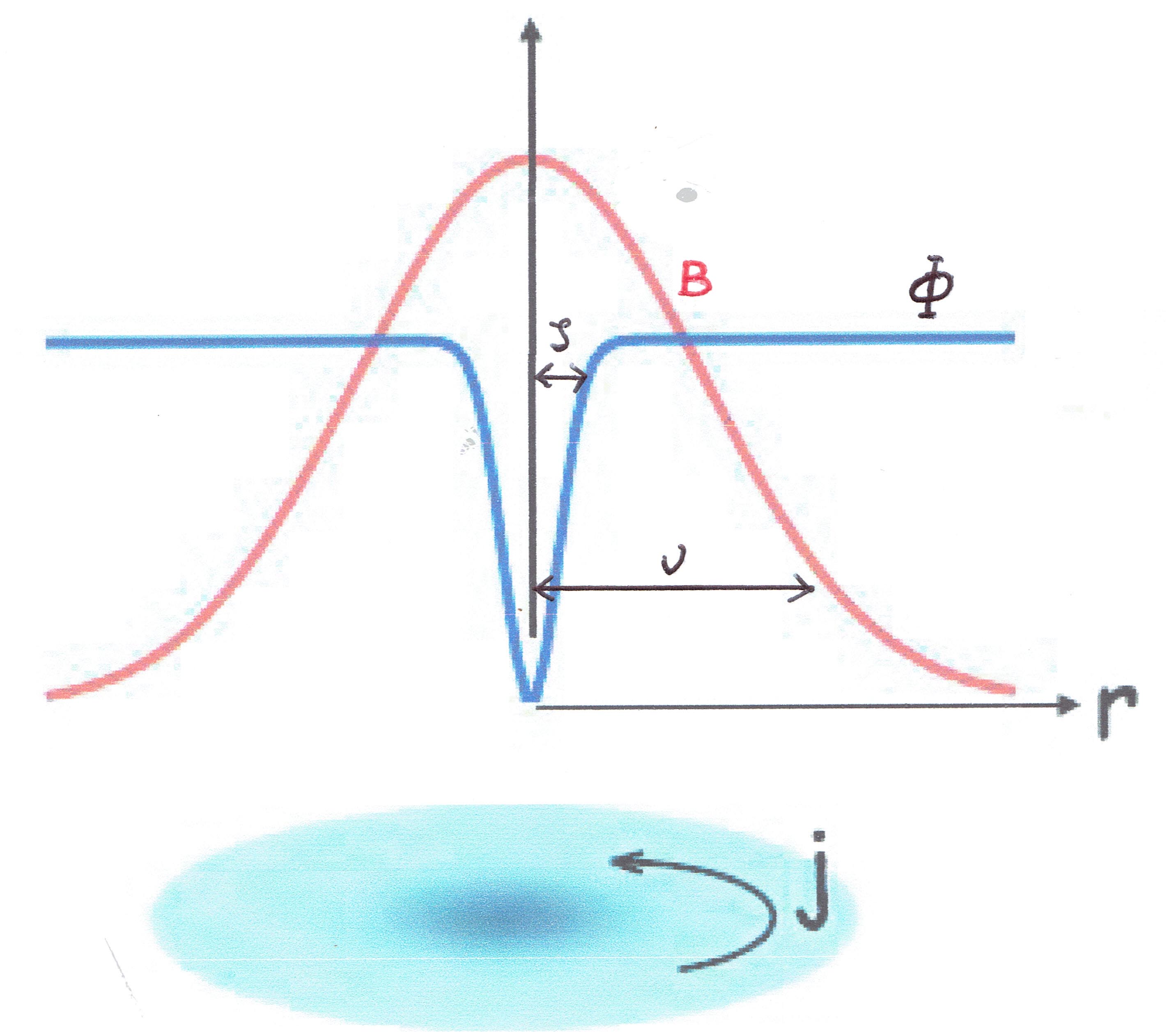}
\includegraphics[width=6.5cm]{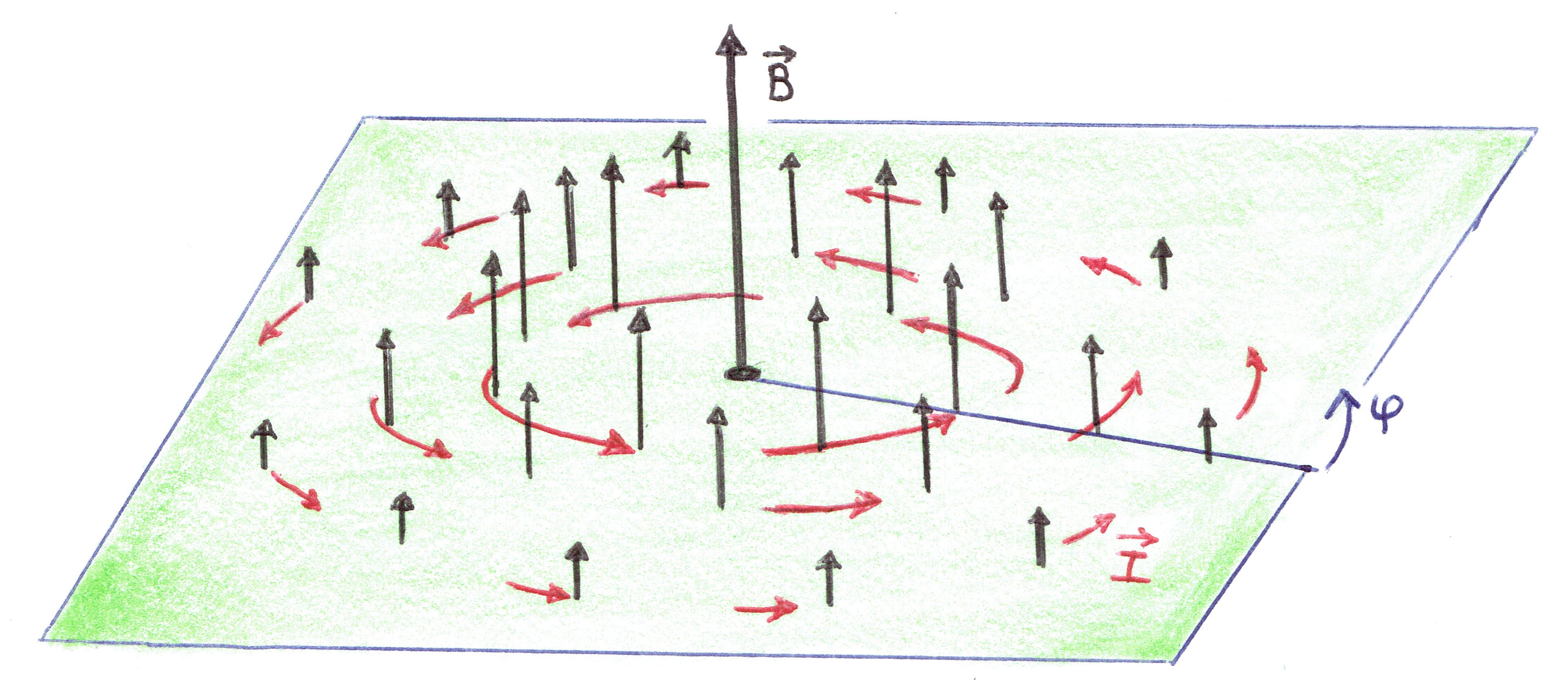}}
\vskip 1.2cm
\caption{{\it Abrikosov (Nielsen-Olesen) vortex in mixed state of quantized flux lines. Vortex supercurrents are sketched by round arrows in red. The radial dependence of the order parameter $\Psi$ is sketched as well as the magnetic field B. In type II superconductivity the coherence length ($\zeta$) is much smaller than the penetration length ($\nu$)}}.
\end{figure}
By the assumption of Eq.(\ref{eqn2}), we see that the Higgs field has a non trivial winding number $n$ and represents de phase jump of $2\pi n$ when the Higgs field makes  a closed curve around the string-like configuration.
When a configuration carries multiple flux quanta $n$, then it was found that the static mass per unit length of the vortex string is given by
\begin{equation}
\mu_n =\int \sqrt{{^2}g}T^{0}_{0}drd\varphi =2\pi\eta^2[n+f(n)]\label{eqn20},
\end{equation}
with $f(n)$ an expression in $n$ via the core radii of the scalar and gauge field, i.e., $r_\Phi\approx\frac{n^\chi}{\sqrt{\beta}\eta}, r_A\approx\frac{n^\tau}{\epsilon\eta}$ ($\tau , \chi$ some constants).
So the question is if an $n$-vortex system will be stable when $n$ grows. The system is not necessarily stable against dissociation into $n$ unit vortices, because the strengths of the electro-magnetic repulsive and  the scalar attractive forces depend on the ratio of scalar to gauge field masses $\alpha \equiv\sqrt{\frac{m_A}{m_\Phi}}=\frac{\epsilon^2}{\beta}$, which can change from a value $\alpha >1$ into $\alpha < 1$. In the special case $\alpha =1$ there are for any separation static solutions.
In the time dependent case, there is a gradient flow and it is conjectured that for winding number $n>1$ there will not generically be uniform convergence due to the escape of vortices to infinity. However, on a compact
set convergence may still persist, as we shall see, in the case when gravity comes into play.
In general, the multi-vortex solution on a time-dependent setting, is governed by highly nonlinear PDE's and is as such  a complicated issue.

One finds for the several orders of the energy density
\begin{eqnarray}
{^{4}{\bar T}}_{tt}=\frac{e^{2\bar\psi}}{2\bar W_1^2 r^2\epsilon^2}(\partial_r\bar P^2+\partial_t\bar P^2)
+\frac{1}{2}(\partial_t\bar X^2+\partial_r\bar X^2)
+\frac{1}{2r^2}e^{2\bar\gamma}\bar X^2\bar P^2+\frac{1}{8}e^{2\bar\gamma-2\bar\psi}\bar W_1^2\beta(\bar X^2-\eta^2)^2,\label{eqn21}
\end{eqnarray}
\begin{eqnarray}
{^{4}T}_{tt}^{(0)}=\dot Y^2+\dot Y(\partial_t\bar X+\partial_r \bar X){\bf cos} [(n_2-n_1)\varphi ]
+\frac{e^{2\bar\psi}}{\bar W_1^2 r^2 \epsilon}\Bigr(\epsilon \dot B^2+\dot B(\partial_r\bar P+\partial_t\bar P)\Bigr),\label{eqn22}
\end{eqnarray}
\begin{eqnarray}
{^{4}T}_{tt}^{(1)}=\dot Z(\partial_t\bar X+\partial_r\bar X){\bf cos}[(n_3-n_1)\varphi]+2\dot Y\dot Z{\bf cos}[(n_3-n_2)\varphi]+2\epsilon\bar X\dot Y B_0{\bf sin}[(n_2-n_1)\varphi]\cr
+\bar X Y\Bigl(\frac{\beta}{2}e^{2\bar\gamma-2\bar\psi}\bar W_1^2(\bar X^2-\eta^2)+\frac{\partial_r\bar X\partial_r Y+\partial_t\bar X\partial_t Y}{\bar X Y}
+\frac{e^{2\bar\gamma}}{r^2}(n_2-n_1+\bar P)\Bigr){\bf cos}[(n_2-n_1)\varphi]\cr
-\frac{e^{4\bar\psi}}{\bar W_1^4r^4\epsilon^2}\Bigl(\frac{1}{2}(\partial_t\bar P^2+\partial_r\bar P^2)+\epsilon\dot B(\partial_t\bar P+\partial_r\bar P)
+\epsilon^2\dot B^2+\frac{1}{2}e^{2\bar\gamma-2\bar\psi}\bar X^2\bar P^2\bar W_1^2\epsilon^2\Bigr)h_{44}\cr
-\Bigl(\frac{1}{8}\beta(\bar X^2-\eta^2)^2+\frac{e^{2\bar\psi}}{2\bar W_1^2r^2}\bar X^2\bar P^2\Bigr) h_{11}
+\dot Y(\partial_t Y+\partial_r Y)
+\frac{e^{2\bar\psi}}{\bar W_1^2r^2\epsilon}\dot C(\partial_t\bar P+\partial_r\bar P+2\epsilon \dot B)+\frac{\epsilon e^{2\bar\gamma}}{r^2}\bar X^2\bar P B\label{eqn23}.
\end{eqnarray}
The background contribution, Eq.(\ref{eqn21}), consists of the well known terms, i.e., the gradients of the gauge field, the gradients of the scalar field,  the coupling term and the contribution from the effective potential.
The first and last terms contain the scale factor ${\cal W}_1$ ( of warp factor) in the denominator and numerator respectively. The contribution of the warp factor depends crucially   on the age of the universe.
In the accelerating stage, the contribution causes an  exponentially amplification [\cite{slag1}]. Here we are dealing with the beginning of the radiation dominated period, just after the symmetry breaking of the model where the warp factor has the opposite feature. See figure 2.
\begin{figure}[h]
\centerline{
\includegraphics[width=5.cm]{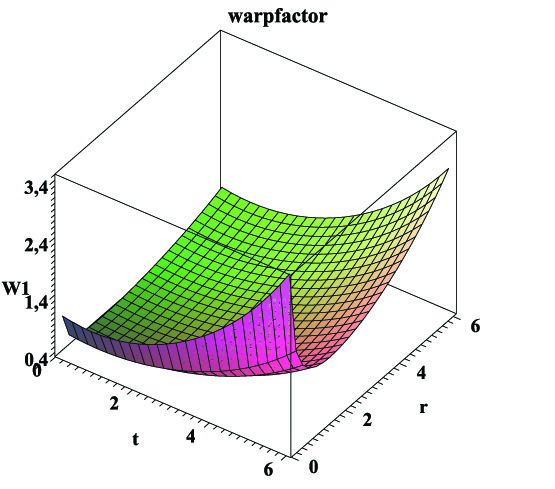}}
\vskip 1.2cm
\caption{{\it The warp factor $W_1(t,r)=\frac{\pm1}{\sqrt{\tau r}} \sqrt{\Bigl(d_1 e^{(\sqrt{2\tau})t}-d_2e^{-(\sqrt{2\tau})t}\Bigr)\Bigl(d_3 e^{(\sqrt{2\tau})r}-d_4e^{-(\sqrt{2\tau})r}\Bigr)}$  plotted for some constants $\tau$ and $d_i$}}.
\end{figure}
Another interesting feature of the model is the behavior of the  $T_{\varphi\varphi}$ components:
\begin{eqnarray}
{^{4}T}_{\varphi\varphi}^{(0)}=e^{-2\gamma}r^2\dot Y(\partial_t\bar X-\partial_r\bar X){\bf cos}[(n_2-n_1)\varphi]+\frac{e^{2\bar\psi-2\bar\gamma}}{\bar W_1^2\epsilon}\dot B(\partial_r\bar P-\partial_t\bar P),\label{eqn24}
\end{eqnarray}
\begin{eqnarray}
{^{4}T}_{\varphi\varphi}^{(1)}=e^{-2\gamma}r^2\dot Z(\partial_t\bar X-\partial_r\bar X){\bf cos}[(n_3-n_1)\varphi]+\frac{e^{2\bar\psi-2\bar\gamma}}{\bar W_1^2\epsilon}\dot C(\partial_r\bar P-\partial_t\bar P)
+e^{-2\bar\gamma}r^2\dot Y(\partial_t Y-\partial_r Y)+\bar X^2\bar P\epsilon B\cr
+\Bigl[\frac{e^{2\bar\psi-2\bar\gamma}}{\bar W_1^2}\dot Y(\partial_t\bar X-\partial_r\bar X)(h_{44}+e^{-2\bar\gamma}r^2h_{11})+\bar X\bar P Y(n_2-n_1+\bar P)+
\frac{1}{2}\beta e^{-2\bar\psi}\bar W_1^2r^2\bar X Y(\eta^2-\bar X^2)\cr
+e^{-2\bar\gamma}r^2(\partial_t\bar X\partial_t Y-\partial_r\bar X\partial_r Y)\Bigr]{\bf cos}[(n_2-n_1)\varphi]
+\frac{e^{4\bar\psi-4\bar\gamma}}{\bar W_1^4r^2\epsilon^2}\Bigl[r^2\epsilon\dot B(\partial_r\bar P-\partial_t\bar P)+\frac{1}{2}r^2(\partial_r\bar P^2-\partial_t\bar P^2)\cr
+\frac{1}{2}\bar W_1^2\epsilon^2e^{2\bar\psi} (\partial_t\bar X^2-\partial_r\bar X^2)\Bigr]h_{11}+\Bigl[\frac{1}{2\bar W_1^2}e^{2\bar\psi-2\bar\gamma}(\partial_t\bar X^2-\partial_r\bar X^2)
-\frac{1}{8}\beta(\bar X^2-\eta^2)^2\Bigr]h_{44}.\label{eqn25}
\end{eqnarray}
The first terms are ${\bf cos}[(n_2-n_1)\varphi]$ and ${\bf cos}[(n_3-n_1)\varphi]$ respectively, indicating that the second-order preferred azimuthal-angle  dependency differs an integer factor.
The alternating behavior between pressure ( positive sign) and tension (negative sign) depends not only on $\alpha$ and the radius of the core of the string as in the static 4D case, but also on the behavior of the warp factor on different time scales.

In our perturbative approximation of  section 2A, we can now construct approximative ground states consisting of correlated separated Nielsen-Olesen vortices.
The conserved charge and current  are given by
\begin{eqnarray}
{\cal I}_\mu=-\frac{1}{2}i\epsilon\Bigl(\Phi(D_\nu\Phi)^*-\Phi^* D_\nu \Phi\Bigr),\quad {\cal Q}=\int \sqrt{-{^3}g}{\cal I}^0drdtd\varphi.\label{eqn26}
\end{eqnarray}
The supercurrents in  strings will increase for  higher winding number.
A string possessing both charge and current densities will have a contribution to the longitudinal momentum $T_{zz}$ and angular momentum ${\cal J}\sim \epsilon_{ij}\int d^2{\bf x}(x^iT^{0j}-x^jT^{0i})$.
Excitations of the vortex lattice will break the axially symmetry. So there is no longer translational symmetry in the z-direction.
This is evident by considering Eq.(\ref{eqn17}) and Eq.(\ref{eqn18}) and by calculating  the first order $(z,z)$ component of the energy momentum tensor
\begin{eqnarray}
{^{4}T}_{zz}^{(0)}=e^{4\bar\psi-2\bar\gamma}\dot Y(\partial_t\bar X-\partial_r \bar X){\bf cos} [(n_2-n_1)\varphi]+\frac{e^{6\bar\psi-2\bar\gamma}}{\bar W_1^2r^2\epsilon}\dot B(\partial_t\bar P-\partial_r\bar P)\label{eqn27}.
\end{eqnarray}
Now the first and second order perturbations of the scalar and gauge fields in higher winding number-mode will decay into NO-strings of lower winding number till the groundstate ( $n=1$) is reached, because it is energetically favourable. See Figure 3. Energy is released by emission of gravitational and EM radiation and the axially symmetry is restored. However the imprint of  the preferred azimuthal-angle is left over in ${^{4}T}_{zz}^{(0)}$ [ and more complicated in ${^{4}T}_{zz}^{(1)}$].
\begin{figure}[h]
\centerline{
\includegraphics[width=6.cm]{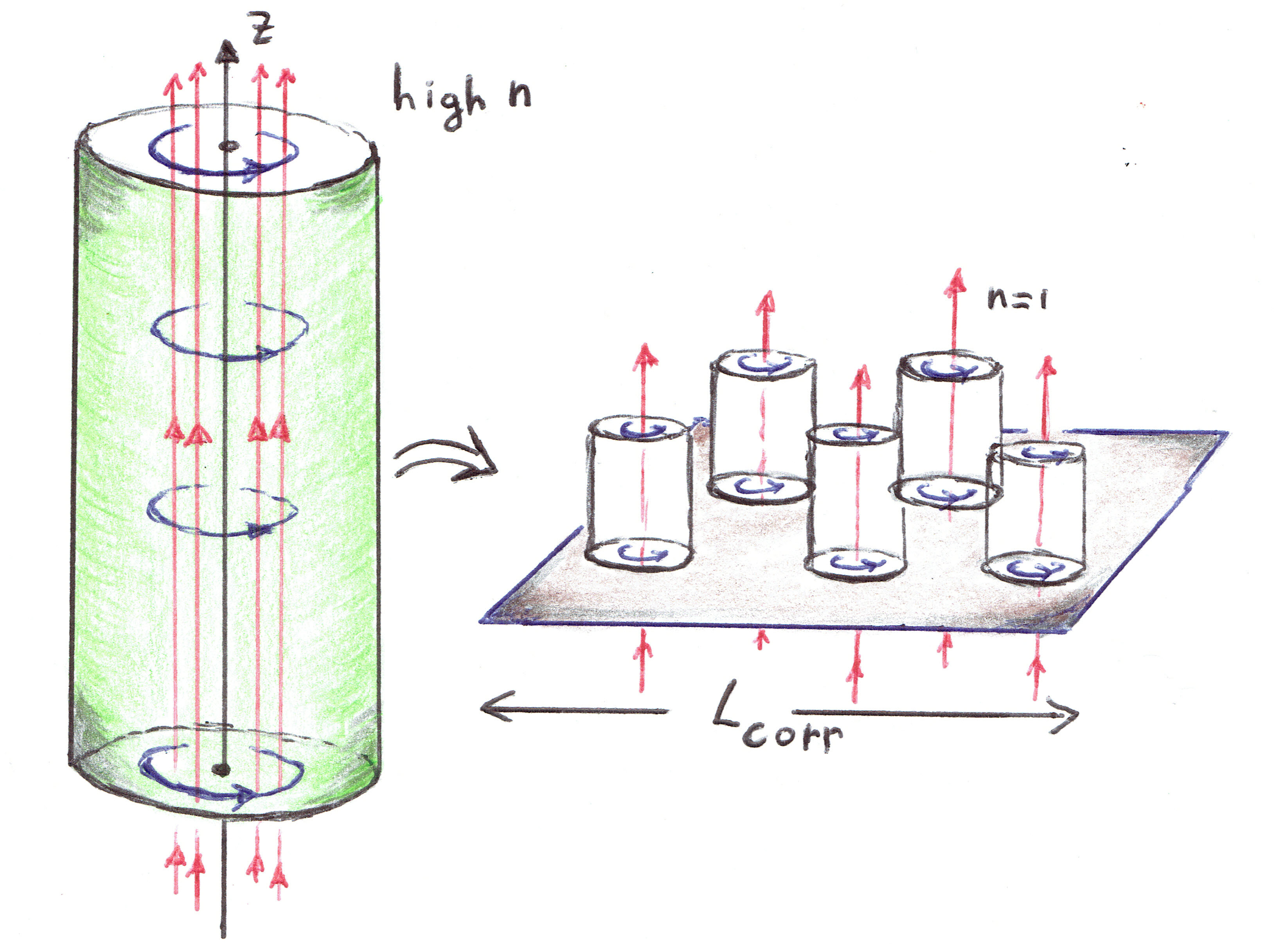}
\includegraphics[width=8.cm]{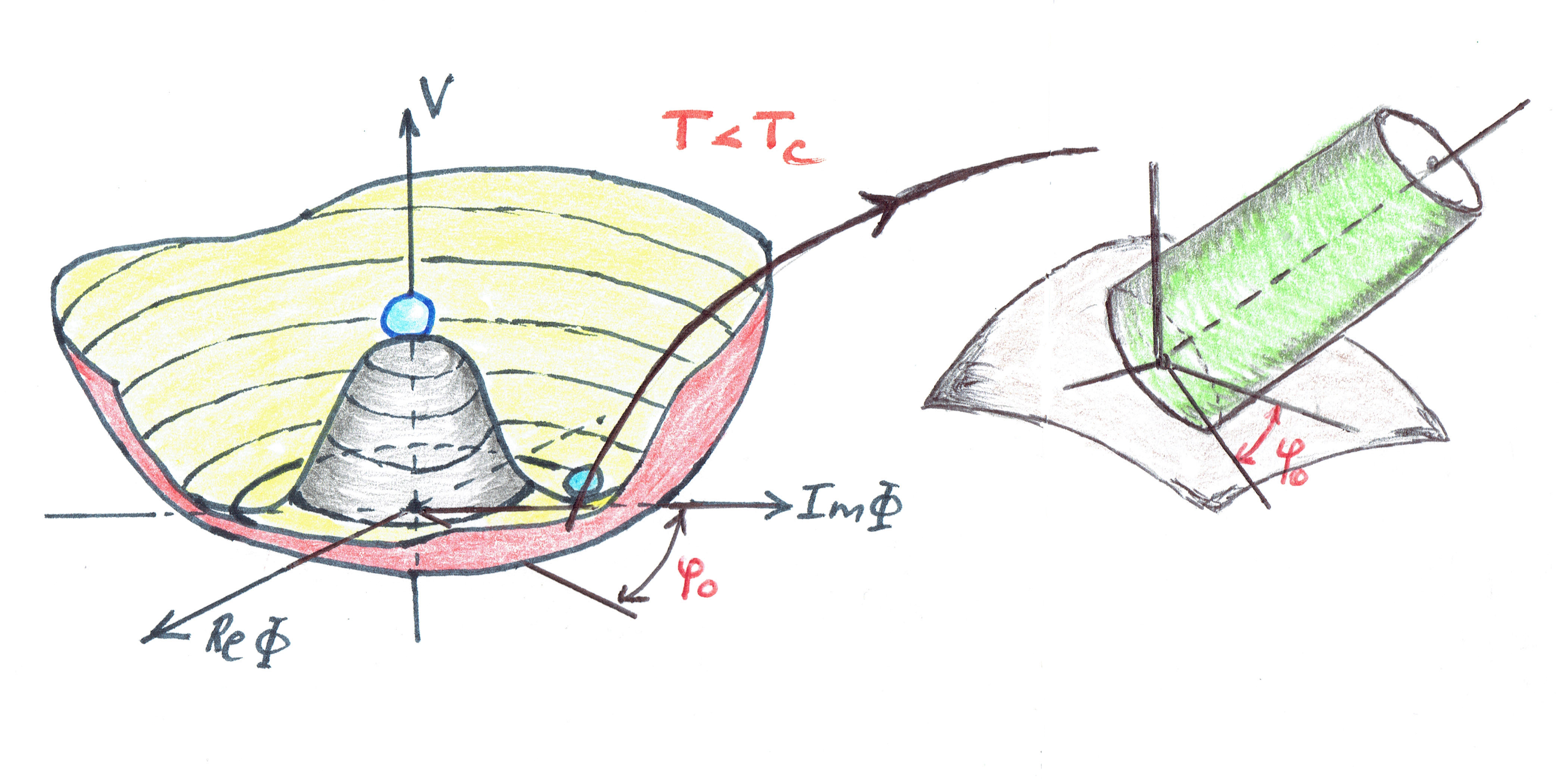}}
\caption{{\it Left: exitation  and decay  of a n-vortex string into correlated ground state Abrikosov vortices ($n=1$)on correlation length $L_{corr}$. Right: the onset of a preferred azimuthal-angle after symmetry breaking.  }}.
\end{figure}
\begin{figure}[h]
\centerline{
\includegraphics[width=3.cm]{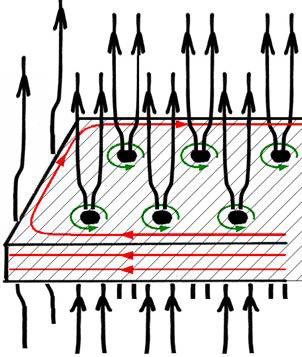}
\includegraphics[width=8.cm]{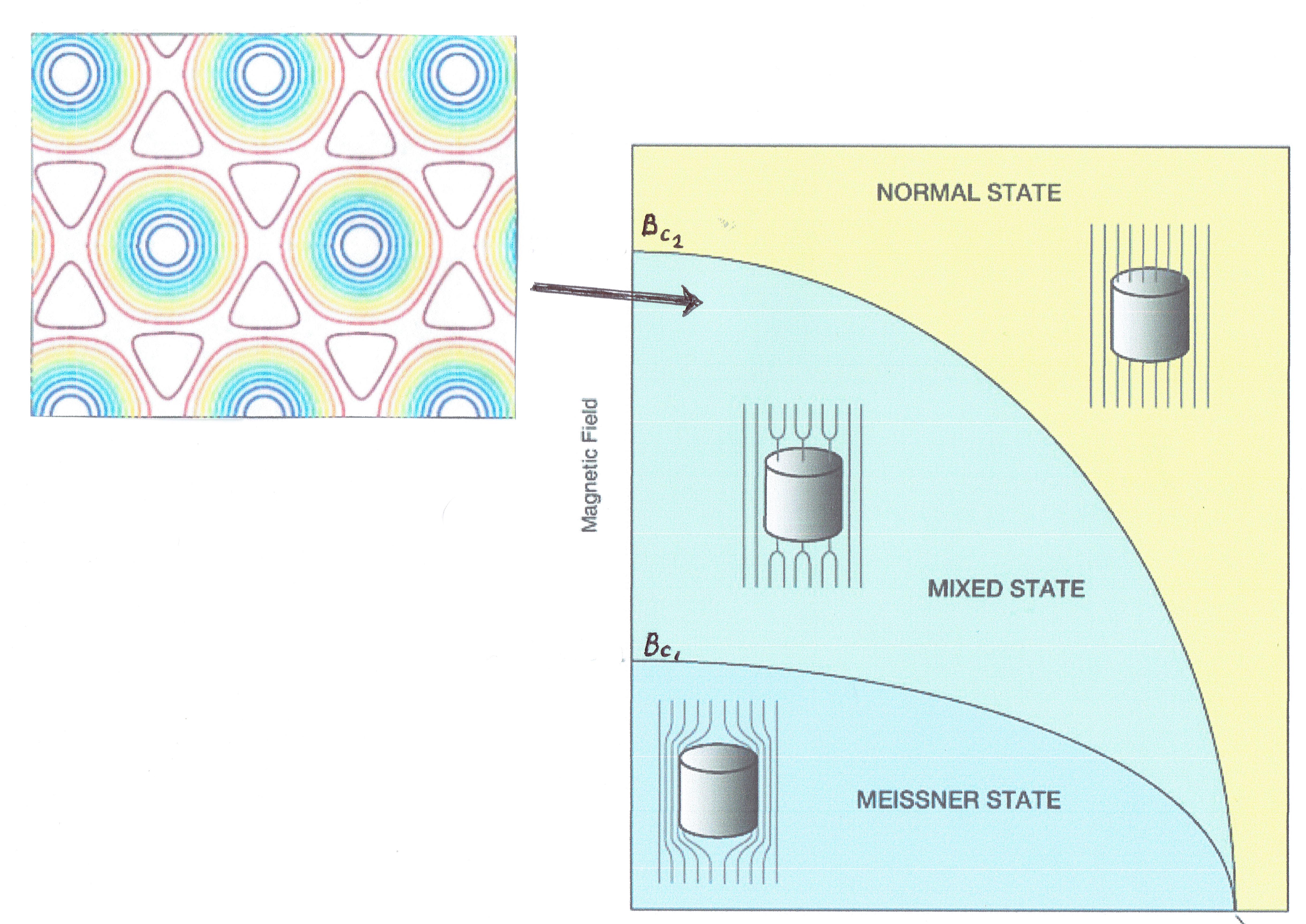}}
\caption{{\it Left: stable hexagonal Abrikosov lattice in mixed state of quantized flux lines. Vortex supercurrents are sketched by round arrows (green). Right: closely packed vortices in the lattice near the critical region  $B_{c_2}$.}}.
\end{figure}
One can prove that the decrease of the supercurrent by decreasing the flux ("phase slip") in the excited string is energetically not preferred: this requires  moving flux from inside to outside the cylinder. This
decrease induces an opposition ( "Lenz-effect")
The resulting periodic vortex lattice carrying a single flux quantum has a hexagonal ( triangular) structure, already predicted by Abrikosov and is experimentally confirmed\cite{abr}. See figure 4. This configuration is stable against perturbations ( "elasticity of the lattice"). It is now conjectured that these localized lattices will carry a common azimuthal-angle preference.

In the next section we will consider the application of our model on the quasar alignment.
\section{\label{sec:level3}The Quasar Link}
\subsection{\label{sec:level3A}Breaking the axial symmetry}
It has been shown\cite{chan,chan1,bert} that self-gravitating  compact objects in equilibrium exhibit the phenomenon of bifurcation along Maclaurin- Jacobi sequences accompanied by spontaneous symmetry breaking similar to the second order phase transition in type II superconductivity. An initial axially symmetric configuration, as is the case in our perturbative model, can dynamically spontaneously be broken, where equatorial eccentricity plays the role of order-parameter. The equatorial eccentricity $\varepsilon\equiv\frac{b}{a}$, with b and a the two equatorial axes, can be expressed through the azimuthal-angle $\varphi (t)$.
The particular orientation of the ellipsoid in the frame $(r,\varphi,z)$ ( see Figure 5) expressed through  $\varphi_0\equiv\varphi(t_0)$, will be at $t>t_0$ determined by the transformation $\varphi \rightarrow \varphi_0- Jt$, where $J$ is the rotation frequency (circulation or "angular momentum") of the coordinate system. The angle $\varphi_0$ is fixed arbitrarily at the onset of symmetry breaking. This arbitrariness of $\varphi_0$, i.e., the orientation of the ellipsoid at $t=t_0$ can be compared with the massless Goldstone-boson modes of the spontaneously broken symmetry of continuous groups.

The phase transition take place on the same time scale  that the vorticity is destroyed by dissipative mechanism and ${\cal J}$ is lost. The end point is a lower energy state that belongs to the Jacobi or Dedekind sequence of equilibrium ellipsoids\cite{chris}.

In the original paper of Chandrasekhar and Lebovitz\cite{chan}, in the Newtonian case, the deformations of the axisymmetric configuration by an infinitesimal nonaxisymmetric deformation is described  in terms of a Lagrangian displacement $\varsigma^a (r,z,\varphi)=\bar\varsigma^a(r,z)e^{in\varphi}$, with $n$ an integer. However, the real part  of the $e^{in\varphi}$ must be put in by hand, in contrast to our result: it appears in a perturbative way as a first and second order effect.
\begin{figure}[h]
\centerline{
\includegraphics[width=6.cm]{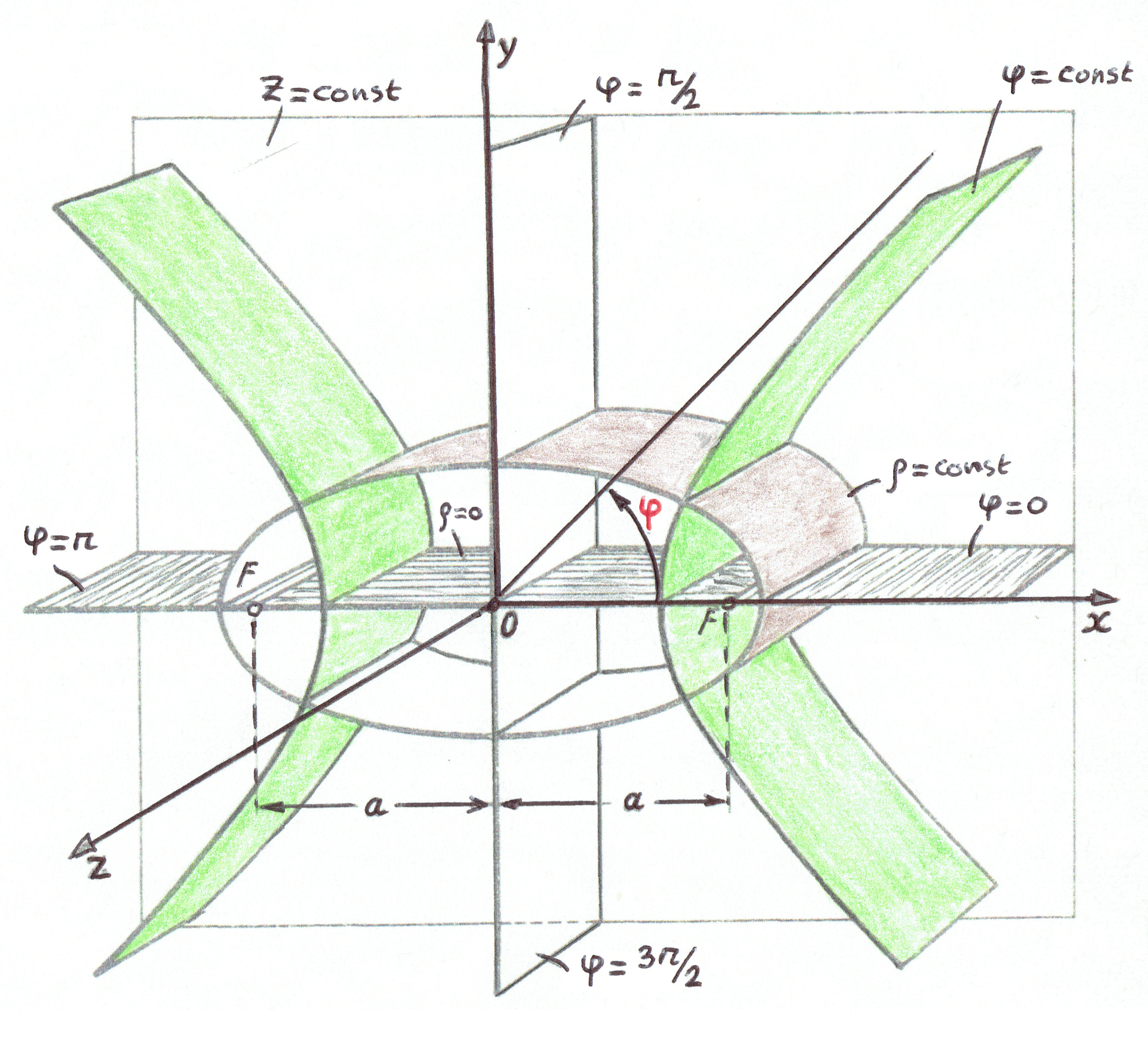}
\includegraphics[width=5.cm]{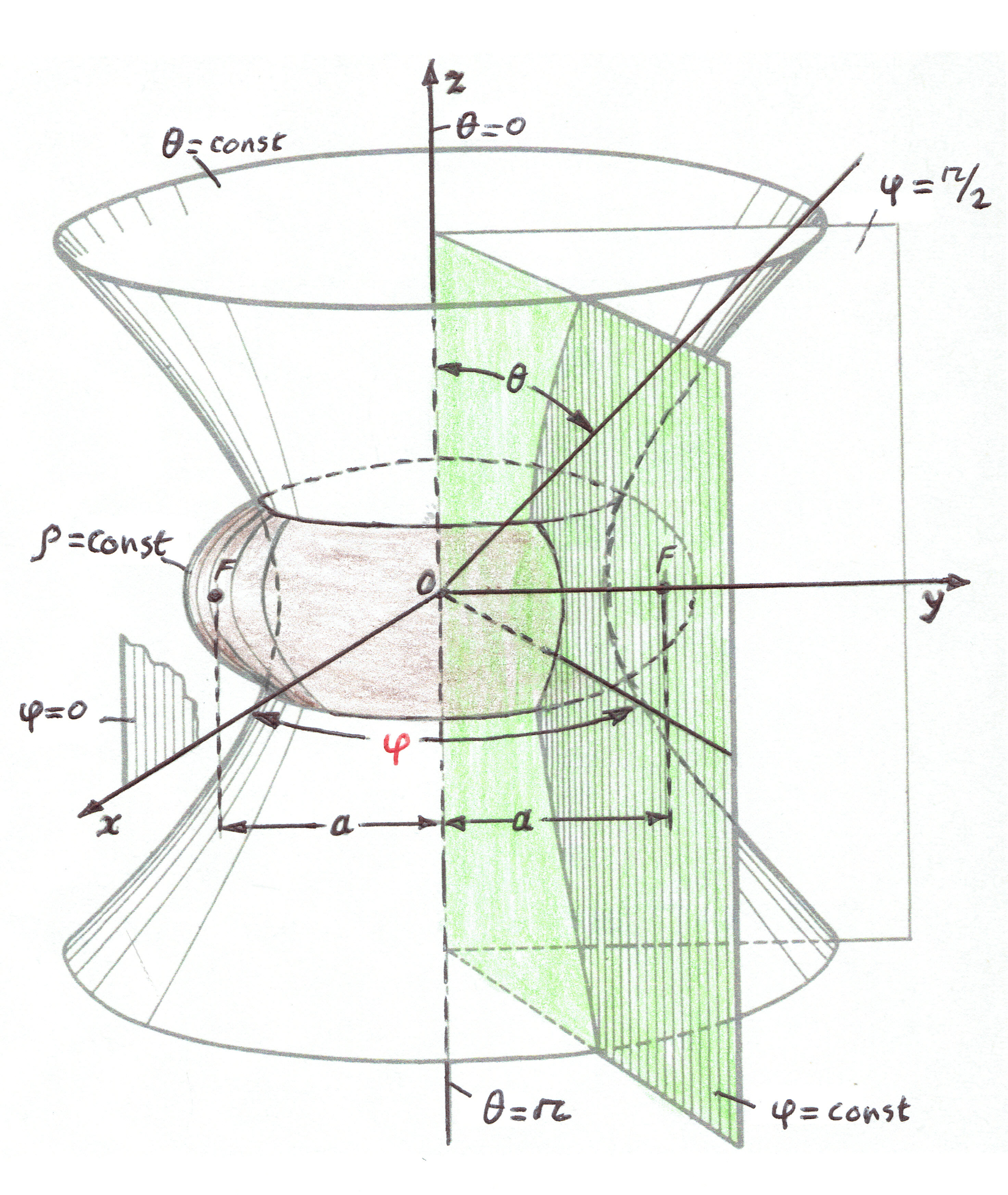}}
\caption{{\it  Left: elliptic-cylindrical coordinates $x=a \cosh \rho\cos\varphi,  y=a \sinh \rho\sin\varphi$ after the temporal disturbances. Right: after transformation to oblate spheroidal coordinates}}.
\end{figure}
The temporarily broken axial symmetry will be the onset of emission of electro-magnetic and gravitational waves, while the string relaxes to the NO configuration.
It is a consequence of the coupled system of PDE's that a high-frequency scalar field can create through an electro-magnetic field, a high frequency gravitational field and conversely.  It is the appearance of the $sin(n_2-n_1)\varphi$ in the first order ${^{4}T}_{t\varphi}^{(0)}$, Eq.(\ref{eqn17}), which triggers this angular momentum and the axially symmetry will be restored when $n_2$ becomes equal to $n_1$ again.
The second order contribution  Eq.(\ref{eqn18}) shows a term $\dot Y h_{14}cos(n_2-n_1)\varphi$, indicating the interaction between the high-frequency EM and gravitational waves. It contains the warp factor in the denominator. In the early stages of the universe $W_1$ is still small and the term $cos(n_2-n_1)\varphi$ is significant and it has a phase difference of $\frac{\pi}{2}$ with respect to the first  order term of Eq.(\ref{eqn17}).
As time increases, it will fade away.
\subsection{\label{sec:level3B}String Evolution, Scaling and the Alignment of Quasar Polarization}
The formation of a network of cosmic strings on large scales when the universe cools down, is often numerically investigated\cite{vil}.
In these 4D models, one usually adopt the Kibble mechanism.  When growing uncorrelated  regions with assigned scaler field phase values meet each other at the boundaries, there will be discrete jumps in the field values.
As the patches with true vacua merge, false vacuum regions are squeezed and form the cosmic strings. The width of the string is roughly $\frac{1}{m_\Phi} \sim \frac{1}{\sqrt{\beta}\eta}$.
These cosmic strings will not be in conflict with standard observational cosmology, in contrast with other topological defects, such as monopoles and domain walls, because the network of strings looses energy by the formation of loops. These loops are chopped off from the long strings by self intersection, starts to oscillate and decay by emitting gravitational and EM energy. The result is that their contribution to $\frac{\rho}{\rho_{crit}}$ remains of the order $G\mu \sim10^{-6}$. So there will be no conflict with the observed acoustic oscillations in the CMB angular power spectrum.
The string evolution is described as "scaled" or scale-invariant, that is, the properties of the network look the same at any particular time  if they are scaled ( or multiplied) by a change in time.
The characteristic scale L of the long string network remains constant relative to $d_H$. The density can be approximated by $\rho_{str}\equiv\frac{\mu}{L^2}$, with $\mu$ the mass per unit length of the string.
This self-similar evolution is confirmed by numerical investigations. All simulations show an evolution to a stable fixed point, where $\rho_{str}t^2=const$\cite{vil}.

Here we adopt  a slightly different scenario. As explained  in section 2B, shortly after the symmetry breaking, a lattice of correlated $n=1$ vortices with a preferred azimuthal-angle, emerged on a correlation length  $L_{corr} <d_H\sim t $. The second order phase transition takes place at the Ginsberg temperature $T_G$ and one finds that $L_{corr}(T_G)\sim\frac{1}{\beta\eta}$. The horizon size at this temperature is $d_H\sim\frac{m_{pl}}{T_G^2}$. So we have $\frac{L_{corr}}{d_H}\sim\frac{\eta}{\beta m_{pl}}$.
These correlated regions will survive to later times, because  at this moment the gravity contribution from the 5D bulk comes into play. The warp factor\cite{slag3} ( see Figure 2)
will have different contributions to the field equations for different times.
The mass per unit length will  contain the warp factor. Just after the symmetry breaking, the vortex will acquire a huge mass $G\mu > 1$ and will initiate the perturbations of high-frequency and justifies our high-frequency approximation.
This is the reason that the regions with $(n=1, \varphi =\varphi_0)$ will stick together and are observed in LQG's with aligned polarization axes\cite{huts,tay}. This alignment of the polarization axes in large quasar groups is observed in the optical range as well in the radio range and cannot be explained by density perturbations.
A side effect of our model is that the alignment of the strings is subgroups of different $\varphi_0$ is probably confirmed by observarion\cite{huts,park}: the aligment of the quasar spin axes is better if the quasar group is divided into smaller systems.
Further, there is observational evidence that in rich LQG's the spin axes of the quasars are preferentially parallel to the major axes of their host LQG, while the spin axes can become also perpendicular to the LQG major axes
when the richness decreases. This can be explained in our model as a second order effect: the higher multiplicity terms, for example in Eq.(\ref{eqn23}) $cos(n_3-n_1)$.
It would be of interest if our second-order alignment effect can be observed if  more data becomes available, specially for high redshift.
If there is a scale-invariant evolution of the network, then one should observe already at high redshift  the vortex clusters predicted in our model.
This will  deliver the proof of not only the cosmological origin of the alignment at a times just after the symmetry breaking, but also new physics beyond standard model such as the extra dimension and self acceleration of the universe without the controversial cosmological constant\cite{slag1,roy3}.
\section{\label{sec:level4}Conclusions}
We find a emergent azimuthal-angle dependency of the Nielsen-Olesen vortices in the general relativistic situation just after the symmetry breaking at GUT-scale.
Using a high-frequency perturbation method, we obtain in the first and second order perturbation equations  $\varphi$-dependent terms left over after the phase transition of the Higgs field.
Vortices  with high multiplicity decay into a lattice with entangled Abrikosov vortices.
The stability of this lattice of correlated flux $n=1$ vortices with preferred azimuthal-angle is  guaranteed by the contribution from the bulk spacetime by means of the warp factor: the cosmic string becomes super-massive for some time during the evolution and initiates the excitations of the vortices to high multiplicity. The correlation will not fade away during the expansion by the warp factor.
We used this azimuthal-angle correlation  for the explanation of the recently observed alignment of polarization axes of quasars in large quasar groups. The detailed behavior of this alignment can be explained with our model. The  two different orientations perpendicular to each other in quasars groups of less richness could be a second order effect in our model.
There is a striking similarity between this  phase transition of the gauged Higgs field  and the temporarily  breaking of the axially symmetry of self-gravitating cosmic string, by the appearance of non-diagonal energy-momentum tensor components. The eccentricity of the ellipsoid can be seen as order parameter. Recovery to SO(2) symmetry induces emission of gravitational and electro-magnetic radiation.
More data of high-redshift quasars will be needed in order to test the second order effect predicted in our model.

\end{document}